\documentclass[pdftex]{pasj01}
\draft

\usepackage{multirow}
\usepackage{color}
\usepackage[dvips]{graphicx}
\usepackage{pslatex} 
\begin{document} 
\Received{}
\Accepted{}

\title{Detailed CO ($J$ = 1--0, 2--1 and 3--2) observations toward an H{\sc ii} region RCW~32 in the Vela Molecular Ridge}

\author{Rei \textsc{enokiya}\altaffilmark{1}, Hidetoshi \textsc{sano}\altaffilmark{1,2}, Katsuhiro \textsc{hayashi}\altaffilmark{1}, Kengo \textsc{tachihara}\altaffilmark{1}, Kazufumi \textsc{torii}\altaffilmark{3}, Hiroaki \textsc{yamamoto}\altaffilmark{1}, Yusuke \textsc{hattori}\altaffilmark{1}, Yutaka \textsc{hasegawa}\altaffilmark{4, 5}, Akio \textsc{ohama}\altaffilmark{1}, Kimihiro \textsc{kimura}\altaffilmark{4}, Hideo \textsc{ogawa}\altaffilmark{4} and Yasuo \textsc{fukui}\altaffilmark{1}}
\altaffiltext{1}{Department of Physics, Nagoya University, Chikusa-ku, Nagoya, Aichi 464-8601, Japan}
\altaffiltext{2}{Institute for Advanced Research, Nagoya University, Furo-cho, Chikusa-ku, Nagoya, Aichi, 464-8601, Japan}
\altaffiltext{3}{Nobeyama Radio Observatory, Minamimaki-mura, Minamisaku-gun, Nagano, 384-1305, Japan}
\altaffiltext{4}{Department of Physical Science, Graduate School of Science, Osaka Prefecture University, 1-1 Gakuen-cho, Naka-ku, Sakai, Osaka 599-8531, Japan}
\altaffiltext{5}{Institute of Space and Astronomical Science, Japan Aerospace Explanation Agency, 3-1-1 Yoshinodai, Chuo-ku, Sagamihara, Kanagawa, 252-5210, Japan}\email{enokiya@a.phys.nagoya-u.ac.jp}




\maketitle


\begin{abstract}
We made CO ($J$ = 1--0, 2--1, and 3--2) observations toward an H{\sc ii} region RCW~32 in the Vela Molecular Ridge. The CO gas distribution associated with the H{\sc ii} region was revealed for the first time at a high resolution of 22 arcsec. The results revealed three distinct velocity components which show correspondence with the optical dark lanes and/or H$\alpha$ distribution. Two of the components show complementary spatial distribution which suggests collisional interaction between them at a relative velocity of $\sim$4 km~s$^{-1}$. Based on these results, we present a hypothesis that cloud-cloud collision determined the cloud distribution and triggered formation of the exciting star ionizing RCW~32. The collision time scale is estimated from the cloud size and the velocity separation to be $\sim$2 Myrs and the collision terminated $\sim$1 Myr ago, which is consistent with an age of the exciting star and the associated cluster. By combing the previous works on the H{\sc ii} regions in the Vela Molecular Ridge, we argue that the majority, at least four, of the H{\sc ii} regions in the Ridge were formed by triggering of cloud-cloud collision.
\end{abstract}

\section{Introduction}
The Vela Molecular Ridge (VMR) was observed for the first time in the $^{12}$CO ($J$ = 1--0) emission line by May et al. (1988). They found a huge molecular complex at distance of 0.4 -- 2.0 kpc toward the Vela region and named it the Vela Molecular Ridge. This complex has total molecular mass of $\sim$ 10$^5$ $M_\odot$ and is divided into four regions the VMR A, B, C and D by Murphy $\&$ May (1991). These authors compared stars whose distances were measured through photometric observations and concluded that the VMR A, C and D are located at 0.7 -- 1.0 kpc while the VMR B at $\sim$ 2.0 kpc. These distances agree with those derived from near infrared reddening by Liseau et al. (1992). 

\subsection{RCW~32}
RCW~32 (Rodgers et al. 1960) also known as Gum 15 (Gum 1955) is the brightest H{\sc ii} region in the VMR D located in the north of the Vela SNR. It looks like the Trifid nebula (M20) at optical wavelengths because of the thin dark lanes. In its vicinity the largest H{\sc ii} region in the VMR D, RCW~27 having a diameter of $\sim$1.7 degrees is located (Figure 1). One of the dark lanes named DC261.5+0.9 or SL2 (Sandqvist \& Lindroos 1976) shows a heliocentric velocity of 22.4 km~s$^{-1}$ corresponding to $V_{LSR}$ = +6.3 km~s$^{-1}$ as measured by CO ($J$ = 2--1), which is consistent with the velocity of the exciting star (Brand et al.1984). The exciting star is suggested to be HD~74804 (HIP~42908) of B0V or B4II (Petteresson $\&$ Reipurth 1994 and references therein) which is located toward the center of a young star cluster Collinder 197 (Cr 197; Collinder 1931). Cr 197 with $\sim$660 $M_{\odot}$ contains a few T--Tauri stars and 75 $\%$ of the members are pre-main sequence stars. The cluster is considered to be an on-going star forming site with an age of $\sim$ 1 Myr (Pettersson $\&$ Reipurth 1994) or 5 $\pm$ 4 Myr (Bonatto $\&$ Bica 2010). UBV photometrical observations (Fitzgerald et al. 1979) and a ZAMS-fit of its cluster members (Vogt $\&$ Moffat 1973) were made and distance of Cr 197 was measured to be 1.02 kpc and 1.05 kpc, respectively, by these authors. Spectroscopic observations of HD~74804 derived 0.72 kpc (Crampton $\&$ Fisher 1974) and 0.67 $\pm$ 0.17 kpc (Pinheiro et al. 2010) as a distance. According to Pettersson $\&$ Reipurth (1994) the different distance between Cr 197 and HD~74804 could be explained if (1) HD~74804 is more luminous than Crampton $\&$ Fisher (1974) assumed, (2) HD~74804 is not a member of Cr 197 or (3) HD~74804 is a double system. Pettersson $\&$ Reipurth (1994) pointed out that radial velocity of HD~74804 may be variable and suggested that the star may be a double system. We consider HD~74804 as the member of Cr 197 and adopt a distance of 1.0 kpc for RCW~32 in the following. The cluster possibly has another early B2V star CD$-$40 4579 which is located at 1.5 pc from its center. Although Hron et al. (1985) assigned CD$-$40 4579 as a member of Cr 197, the heliocentric velocity of CD$-$40 4579, 96.5 km~s$^{-1}$, is significantly different from that of the other member stars, making its membership to Cr 197 ambiguous. Yamaguchi et al. (1999) conducted $^{13}$CO ($J$ = 1--0) observations toward 23 southern H{\sc ii} regions with bright-rimmed clouds (BRCs) and showed that distribution of molecular clouds facing H{\sc ii} regions is affected by UV radiation and forming massive stars possibly via radiation driven implosion (RDI). RCW~32 is accompanied by two BRCs SFO 57 and SFO 58 as marked in Figure 2b (Sugitani $\&$ Ogura 1994). SFO 58 shows a V-shaped appearance characteristic to bright rimmed clouds and is classified as an ``A type'' BRC by Sugitani et al. (1991). Urquhart et al. (2006) carried out $^{12}$CO ($J$ = 1--0), $^{13}$CO ($J$ = 1--0) and C$^{18}$O ($J$ = 1--0) observations toward SFO 58 by the Mopra 22 m telescope and 3.6 cm and 6 cm radio continuum observations by the Australia telescope Compact Array (ATCA), and found that a $^{12}$CO ($J$ = 1--0) cloud, which is located in the bright rim, accompanies an embedded ultracompact H{\sc ii} region excited by a single B2 -- B3 star and possible molecular outflow. The authors concluded that SFO 58 is strongly affected by HD~74804 and the ongoing star formation was triggered by RDI of HD~74804. 

In the present paper, we report results of first multi-$J$ transition CO line observations toward RCW~32 by the NANTEN2, Mopra and Atacama Sub-mm Telescope Experiment (ASTE) telescopes with resolutions of 22$''$ -- 180$''$  and discuss its  formation mechanism by comparing distributions of CO and at infrared and optical wavelengths. The direction in this paper is given in the Galactic coordinate. In Section 2, we describe the observations. Section 3 consists of four subsections; Subsections 3.1 and 3.2 present large- and small-scale gas distribution toward RCW~32 and Subsections 3.3 and 3.4, give physical properties of CO gas in RCW~32. In Section 4, we discuss cloud-cloud collision as a possible formation mechanism of the exciting star of RCW 32 and Section 5 gives conclusions of the present study.

\section{Observations}
\subsection{$^{12}$CO ($J$ = 1--0)  and $^{13}$CO ($J$ = 1--0)}
$^{12}$CO ($J$ = 1--0) and $^{13}$CO ($J$ = 1--0) observations were made by the NANTEN2 4-m mm/sub-mm telescope at Atacama, Chile in October 2011. The half power beam width (HPBW) of the NANTEN2 telescope at $\sim$115 GHz corresponds to $\sim$ 180$''$. The 4 K cooled Nb \textcolor{black}{Superconductor}--Insulator--Superconductor (SIS) double-side band (DSB) mixer receiver installed as the front\textcolor{black}{--}end enabled us to observe both the $^{12}$CO ($J$ = 1--0) and $^{13}$CO ($J$ = 1--0) lines simultaneously. The typical system noise temperatures toward RCW~32 including the atmosphere for $^{12}$CO ($J$ = 1--0) and $^{13}$CO ($J$ = 1--0) were $\sim$200 and $\sim$150 K, respectively. The backend consists of two Acqiris signal analyzers (AC240) as the digital Fourier transform spectrometer (DFS). Each DFS has 16384 channels with a 1 GHz band width and centered on 115.271202 GHz for $^{12}$CO ($J$ = 1--0) and 110.201353 GHz for $^{13}$CO ($J$ = 1--0) provided velocity coverages of $\sim$ 2600 km~s$^{-1}$ and $\sim$ 2700 km~s$^{-1}$ and velocity resolutions of $\sim$ 0.159 km~s$^{-1}$ and $\sim$ 0.166 km~s$^{-1}$, respectively. Observations were conducted by the on-the-fly (OTF) mode with Nyquist Sampling and two orthogonal scan maps (an $x$ scan map and a $y$ scan map) with 1.0 degree by 1.0 degree were summed up to reduce scanning effects. We used the spectrum of ($l$, $b$) = (262\fdg4000, 1\fdg8001), which has no significant emission at 0.25 K (in antenna temperature {\it T$_{a}$$^{*}$}) sensitivity with 0.08 km~s$^{-1}$ velocity resolution, as the reference spectrum for the blank sky. The pointing accuracy was checked every 2 hours and was achieved to be within 10” through cross scan observations of IRC+10216 or the sun. The absolute intensity calibration was done through comparing peak intensity of Orion-KL [$\alpha_\mathrm{J2000}$ = $5^{\mathrm{h}}35^{\mathrm{m}}14\fs48$, $\delta_\mathrm{J2000}$ = $-5{^\circ}22\arcmin27\farcs55$] observed by NANTEN2 (Ridge et al. 2006).

\subsection{$^{12}$CO ($J$ = 2--1)  and $^{13}$CO ($J$ = 2--1) }
$^{12}$CO ($J$ = 2--1) and $^{13}$CO ($J$ = 2--1) observations were made by NANTEN2 from November 4 to 8 in 2015. The HPBW of the NANTEN2 telescope at $\sim$230 GHz corresponds to $\sim$ 90”. The 4 K cooled Nb SIS DSB mixer receiver installed as the front\textcolor{black}{--}end enabled us to observe both $^{12}$CO ($J$ = 2--1) and $^{13}$CO ($J$ = 2--1) lines simultaneously. The typical system noise temperature toward RCW~32 including the atmosphere for the both lines was $\sim$140 K. The backend configuration is the same as CO ($J$ = 1--0) observations but the DFS centered on 230.538000 GHz for $^{12}$CO ($J$ = 2--1) and 220.398681 GHz for $^{13}$CO ($J$ = 2--1) provided velocity coverages of $\sim$ 1300 km~s$^{-1}$ and $\sim$ 1350 km~s$^{-1}$ and velocity resolutions of $\sim$ 0.080 km~s$^{-1}$ and $\sim$ 0.083 km~s$^{-1}$, respectively. Observations were conducted by the OTF mode with Nyquist Sampling and orthogonal two scan maps (an $x$ scan map and a $y$ scan map) with 30 arcmin by 30 arcmin were summed up to reduce scanning effects. We used the spectrum of ($l$, $b$) = (262\fdg3708, 1\fdg2347), which has no significant emission in 0.12 K ({\it T$_{a}$$^{*}$}) sensitivity, as the reference spectrum for the blank sky. The pointing accuracy was checked every 4 hours and was achieved to be better than 10” through cross scan observations of IRC+10216. The absolute intensity calibration was done through pixel-by-pixel comparisons between two OTF maps toward Orion B obtained with NANTEN2 in {\it T$_{a}$$^{*}$} and the 1.85-m telescope by Osaka Prefecture University in main beam temperature, {\it T$_{mb}$} (Onishi et al. 2013; Nishimura et al. 2015). Both maps are convolved with Gaussian to the same beam size 204$\arcsec$ and have a field of 28$\arcmin$ by 28$\arcmin$ consisting of 28 pixels by 28 pixels centered on ($l$, $b$) = (206\fdg45, -16\fdg33) with an integrated velocity range from 1.7 to 15.7 km~s$^{-1}$. Then, the scaling factor which converts {\it T$_{a}$$^{*}$} to {\it T$_{mb}$} for NANTEN2 by comparing NANTEN2 data in {\it T$_{a}$$^{*}$} and 1.85 m data in {\it T$_{mb}$} was derived through a linear fitting between the whole 28 $\times$ 28 pixels of the two maps. The obtained scaling factors and correlation coefficients are 1.277 $\pm$ 0.003 and 0.971 for $^{12}$CO and 1.238 $\pm$ 0.006 and 0.968 for $^{13}$CO, respectively. The peak intensities in the calibrated maps were 149 K~km~s$^{-1}$ for $^{12}$CO ($J$ = 2--1) toward ($l$, $b$) = (206\fdg41, -16\fdg26) and 58 K~km~s$^{-1}$ for $^{13}$CO ($J$ = 2--1) toward ($l$, $b$) = (206\fdg56, -16\fdg36).

\subsection{$^{12}$CO ($J$ = 3--2) }
$^{12}$CO ($J$ = 3--2) datasets were observed by the ASTE 10 m sub-mm telescope at Atacama, Chile from November 29 to December 5 in 2016 (Ezawa et al. 2004; 2008). The HPBW of the ASTE telescope at $\sim$ 345 GHz corresponds to $\sim$ 22$\arcsec$. A cartridge-type dual-polarization side-band separating mixer receiver for the 350 GHz band, ``DASH 345``, was used for the observations. The typical system noise temperature toward RCW~32 including the atmosphere was $\sim$ 300 K. Two XF-type digital spectro-correlators, ``MAC`` with the narrow band mode (bandwidth = 128 MHz) centered on 345.795990 GHz which provided a velocity coverage of $\sim$111 km~s$^{-1}$ and velocity resolution of $\sim$ 0.11 km~s$^{-1}$ were selected for the observations (Sorai et al. 2000). Our observations were conducted by the OTF mode with Nyquist Sampling and orthogonal two scan maps (an $x$ scan map and a $y$ scan map) with 8 $\arcmin$ $\times$ 8 $\arcmin$ were summed up to reduce scanning effects. We used the spectrum of ($l$, $b$) = (262\fdg46, 1\fdg71), which has no significant emission in 0.12 K ({\it T$_{a}$$^{*}$}) sensitivity, as the reference spectrum for the blank sky. The pointing accuracy was checked every 2 hours and was achieved to be within 3$\arcsec$ through cross scan observations of RAFGL 4078 (07h45m02.41s, $-$71d19m45.7s) and GX Mon (06h52m47.04s, +08d25m18.8s). The absolute intensity calibration was done by comparing peak intensity of IRC+10216 [$\alpha_\mathrm{B1950}$ = $9^{\mathrm{h}}45^{\mathrm{m}}14\fs8$, $\delta_\mathrm{B1950}$ = $-13{^\circ}30\arcmin40\arcsec$] observed by the ASTE telescope and the CSO telescope to be 32.5 K (Wang et al. 1994).

\section{Results}
\subsection{Gas Distributions in the Vicinity of RCW~32}
\begin{figure}[htbp]
 \begin{center}
  \includegraphics[width=14cm]{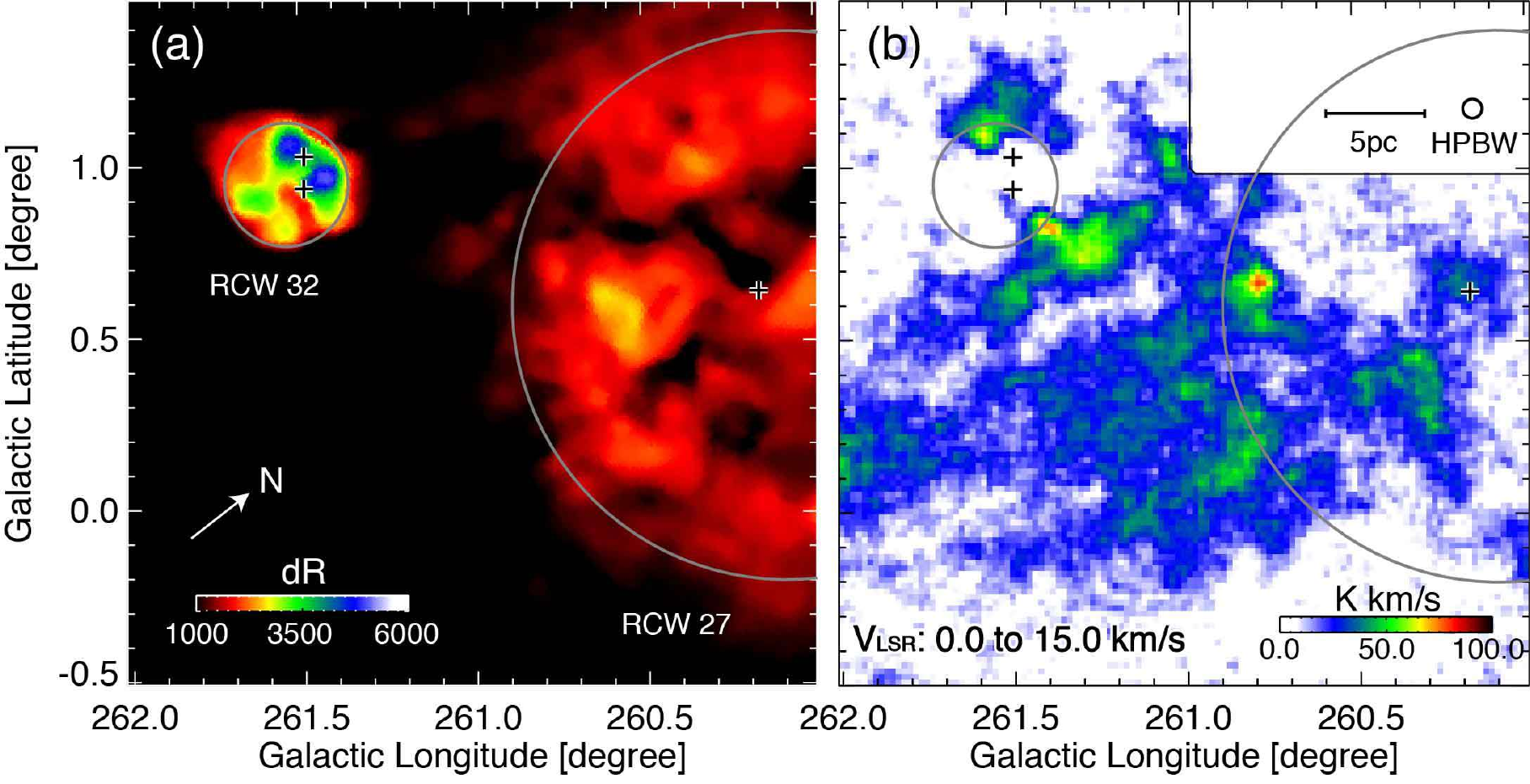} 
 \end{center}
\caption{Large-scale molecular / ionized gas distributions of RCW~32. The small and large gray circles indicate the size of RCW~32 and 27 respectively. The cross in RCW~27 indicates the position of the exciting star and crosses in RCW~32 indicate positions of early B stars toward RCW~32. The arrow indicates the north in the equatorial coordinate for reference. (a) Distribution of the H$\alpha$ recombination line emission obtained by smoothed continuum-subtracted image of the Southern H-Alpha Sky Survey Atlas (SHASSA; Gaustad et al. 2001). The unit of H$\alpha$ is dR (decirayleighs). (b) Integrated intensity distribution of the $^{12}$CO ($J$ = 1--0) with velocity integration range of 0.0 to 15.0 km~s$^{-1}$ obtained by the NANTEN2 telescope.}\label{large}
\end{figure}

Figure~1 shows large-scale distributions of the molecular and ionized gas toward RCW 32 in a 2$\arcdeg$ by 2$\arcdeg$ field. Figure~1a shows distribution of the H$\alpha$ recombination line emission (Gaustad et al. 2001). The small and large gray circles indicate approximate shapes of RCW~32 and RCW~27, respectively. Crosses show positions of early B stars (CD$-$40 4579 and HD~74804) toward RCW~32 and the exciting star (HD~73882) of RCW~27. In H$\alpha$ RCW~32 is more than twice as bright as RCW~27. The elongated depression from the center of RCW~32 to the southwest corresponds the dark lane SL2 (Brand et al.1984). Figure~1b shows velocity integrated intensity distribution of $^{12}$CO ($J$ = 1--0) observed by the NANTEN2 telescope, whose integrated velocity range is from 0.0 to 15.0 km~s$^{-1}$. The molecular gas is extended over the region and RCW~32 seems to be located between two CO peaks at ($l$, $b$) = (261\fdg40, 0\fdg85) and (261\fdg55, 1\fdg05).

\begin{figure}[htbp]
 \begin{center}
  \includegraphics[width=16cm]{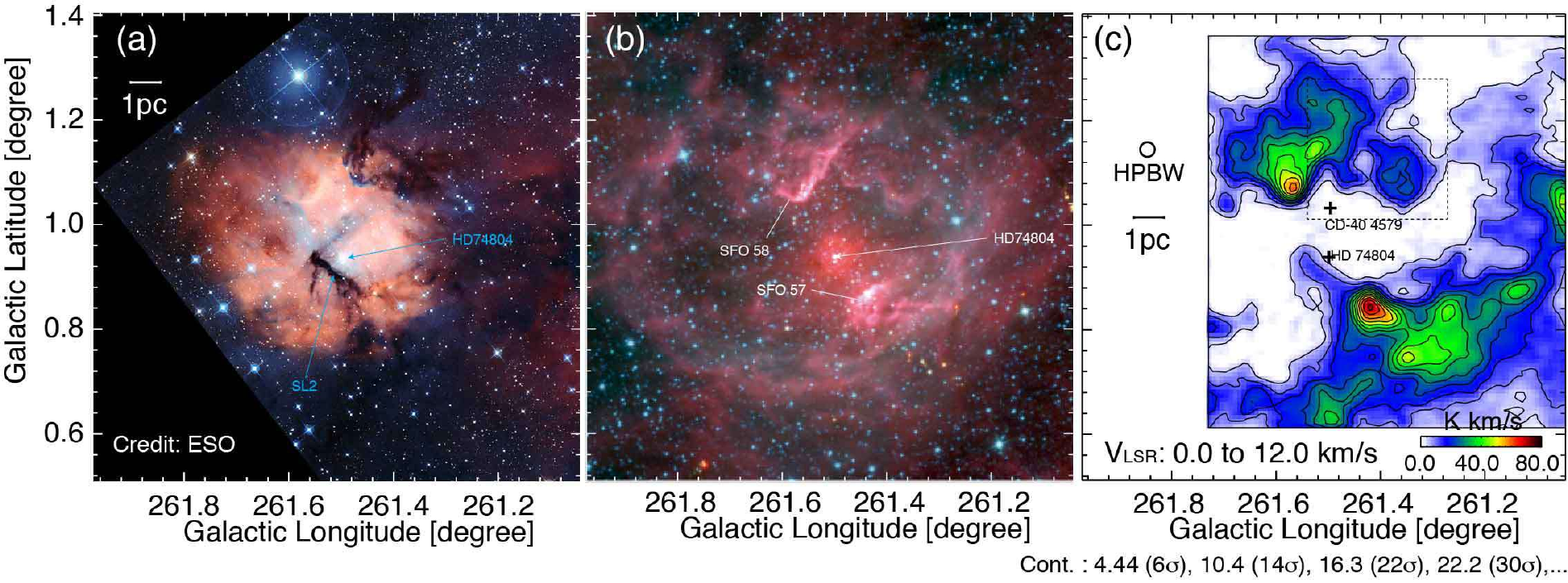} 
 \end{center}
\caption{RCW~32 in various wavelengths. (a) Optical three-color composite image of RCW~32. Blue, yellow, and red correspond to B band, R band, and H$\alpha$ taken by the MPG / ESO 2.2--metre telescope (credit: ESO). (b) Infrared three-color composite image of RCW~32. Blue, green, and red correspond to 3.4 $\mu$m, 4.6 $\mu$m, and 22 $\mu$m obtained by WISE (Wright et al. 2010). (c) Integrated intensity distribution of the $^{12}$CO ($J$ = 2--1) obtained by the NANTEN2 telescope. The dashed square shows the region which we observed in $^{12}$CO ($J$ = 3--2) with the ASTE telescope. Crosses indicate positions of early B stars of CD$-$40 4579 and HD~74804 toward this region.}\label{appearance}
\end{figure}

Figure~2 shows close-up views toward RCW~32 at optical and infrared wavelengths and the CO distribution. Figure~2a shows an optical three composite color image of RCW~32 obtained by Wide Field Imager installed at MPG / ESO 2.2-metre telescope in La Silla, Chile (Credit: ESO\footnote[1]{https://www.eso.org/public/images/eso1420c/}). The red, green, and blue represent H$\alpha$, R band, and B band, respectively. The H{\sc ii} region RCW~32 is roughly spherically with a radius of $\sim$3 pc centered on the exciting star HD~74804, and the star cluster Cr 197 including HD~74804 is extended with a radius of 1.5 pc centered on the star. The dark lane SL2 is seen from the center of RCW~32 to the southwest and another dark cloud is seen toward the northwestern edge of RCW~32. Figure~2b shows an infrared three composite color image\footnote[2]{http://irsa.ipac.caltech.edu/applications/wise/} of RCW~32 obtained with {\it WISE}. The red, green, and blue represent distributions at 22 $\mu$m, 4.6 $\mu$m, and 3.4 $\mu$m, respectively. The figure shows that the dust emission at the infrared wavelengths delineates the outer boundary of RCW~32 in particular toward the western half.

In Figure~2b the two bright features in the north and south correspond to the BRCs SFO~58 and SFO~57. The exciting star HD~74804 is located in the center of RCW~32 between these BRCs. The dark lane SL2 is not seen in the infrared wavelength. Figure~2c shows integrated intensity distribution of $^{12}$CO ($J$ = 2--1) observed with NANTEN2 in a velocity range from 0.0 to 12.0 km~s$^{-1}$. The black crosses indicate the positions of stars earlier than B4. The dashed black box indicates the region where we observed in $^{12}$CO ($J$ = 3--2) with ASTE. Figure~2c shows that the two peaks of the molecular clouds are distributed toward the two BRCs. These BRCs are both considered to be driven by the ultraviolet radiation of HD~74804, and the dense molecular cloud toward SFO 58 exhibits an active on-going young massive star formation as shown by the infrared object IRAS08435-4105 (Urquhart et al. 2006). The distribution of the overall molecular gas implies that RCW~32 is not a young H{\sc ii} region, where molecular gas inside of the H{\sc ii} region is almost fully ionized by the exciting star except for the two BRCs which still hold their original cloud shape at least in part.

\begin{figure}[htbp]
 \begin{center}
  \includegraphics[width=14cm]{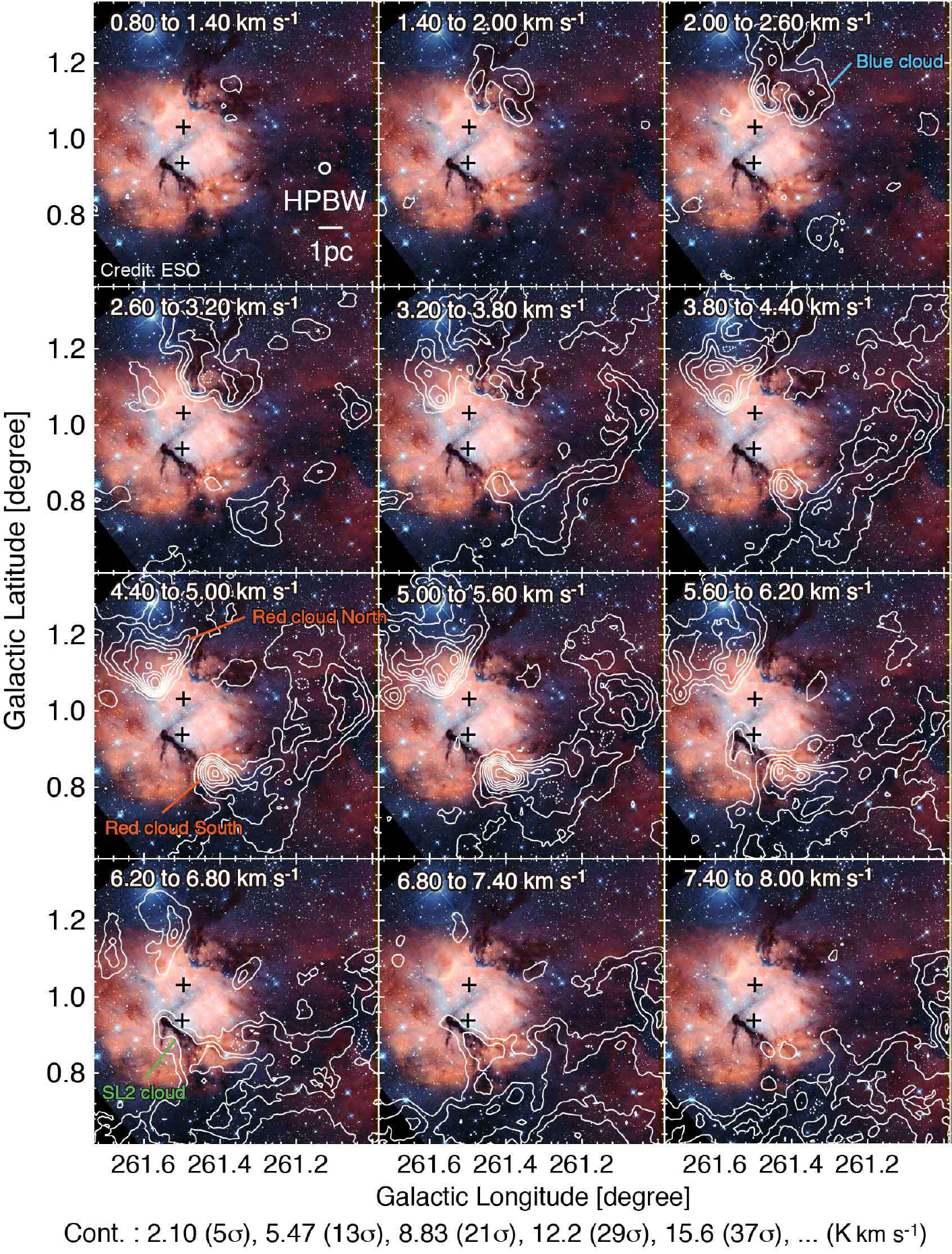} 
 \end{center}
\caption{Velocity channel distribution of $^{12}$CO ($J$ = 2--1) superposed on the optical image of Figure~2a (Credit: ESO) as contours. Contours are from 2.1 K~km~s$^{-1}$ with the step of 3.37 K~km~s$^{-1}$. Velocity integration ranges are shown at the top of each panel. Crosses indicate positions of early B stars in the Figure.}\label{ch21}
\end{figure}

Figure~3 shows velocity-channel distribution of the $^{12}$CO ($J$ = 2--1) emission as contours superposed on the optical composite image of Figure~2a. The black crosses indicate positions of the two early B stars. In a velocity range from 1.4 to 3.8 km~s$^{-1}$, the molecular cloud (hereafter the Blue cloud) corresponds to the dark cloud and the weak molecular emission corresponds to the dark spot in the southeast of RCW~32 around ($l$, $b$) = (261\fdg68, 0\fdg80). In a velocity range from 2.6 to 6.8 km~s$^{-1}$, the molecular cloud (hereafter the Red cloud South) corresponds to SFO 57 and a V-shaped molecular cloud (hereafter the Red cloud North) corresponds to SFO 58. In a velocity range from 3.2 to 5.6 km~s$^{-1}$, a large molecular cloud with a size of $\sim$ 5 pc corresponds to the dark cloud in the west of RCW~32. The cloud shows no strong sign of interaction with the B stars. In a velocity range from 5.6 to 7.4 km~s$^{-1}$, SL2 coincides with $^{12}$CO emission (hereafter the SL2 cloud).

\begin{figure}[htbp]
 \begin{center}
  \includegraphics[width=14cm]{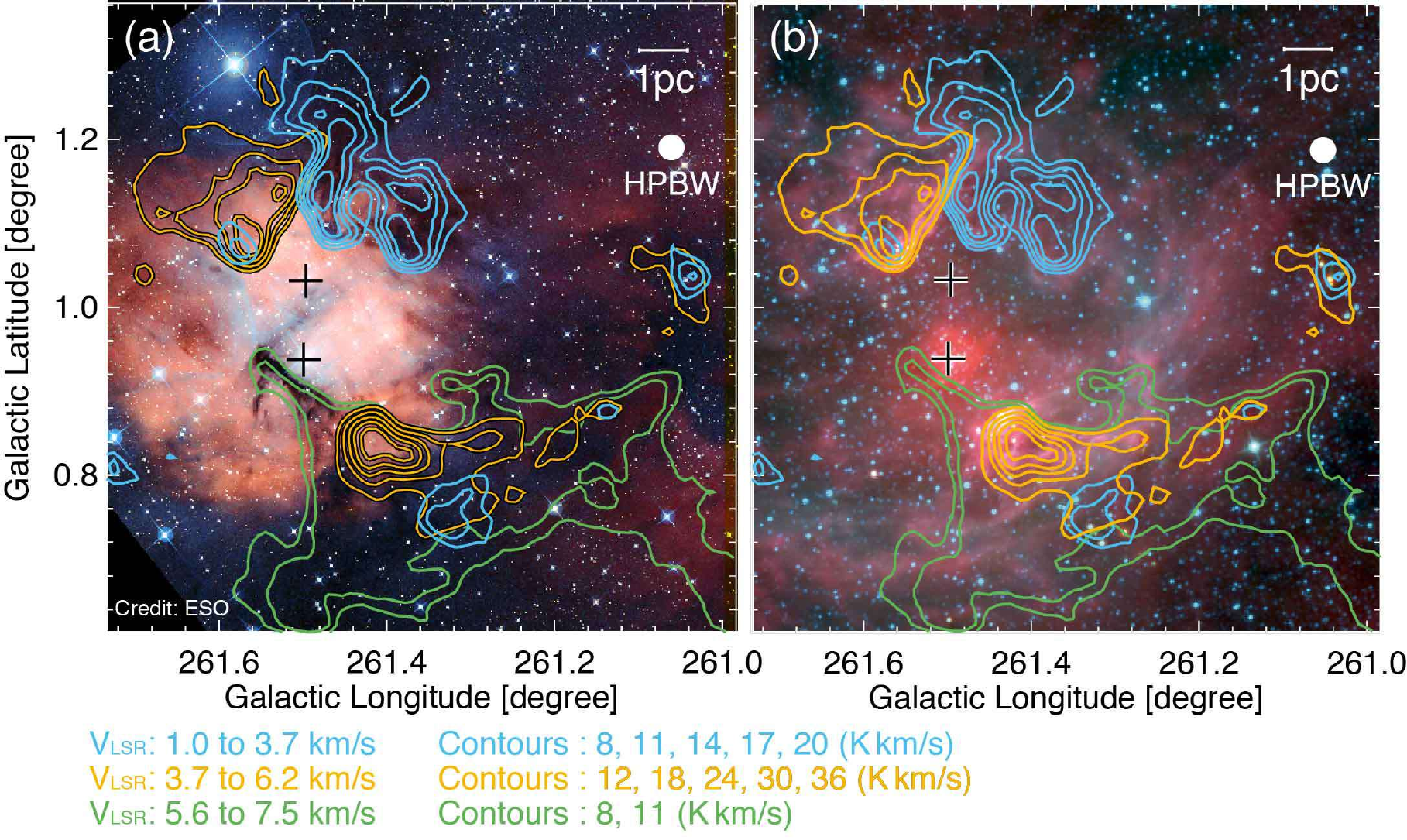} 
 \end{center}
\caption{Integrated intensity distributions of $^{12}$CO ($J$ = 2--1) are superposed on (a) the optical and (b) infrared images as contours. Integrated velocity ranges shown by the orange, blue and green contours correspond to those of the Red, Blue, and SL2 clouds, respectively. Crosses indicate positions of early B stars.}\label{clouds}
\end{figure}

Figure~4 shows $^{12}$CO ($J$ = 2--1) contours of the four clouds (the Red cloud South, the Red cloud North, the Blue cloud, and the SL2 cloud) superposed on a three color composite image as shown in Figures~2a and 2b. Black crosses indicate positions of two early B stars. Figure~4a shows that the two Red clouds are brighter in CO than the other clouds, whereas the SL2 cloud is of the lowest CO intensity. Red clouds North and South are not dark in the optical wavelength or not luminous in the infrared wavelength. This indicates that these two clouds are not on the near side of the H{\sc ii} region RCW~32. On the other hand, the Blue and SL2 clouds are dark in the optical wavelength and should be located in front of RCW~32. The shape of the Blue cloud and its infrared ridge suggests that the cloud is dynamically affected by the B stars. The Blue cloud shows no BRC or enhanced infrared emission toward the B star CD$-$40 4579, but it is apparently close to the B star. We infer that the CD$-$40 4579 is at a significantly larger distance than its projected distance, and the Blue cloud is not physically affected by the star.

\subsection{Detailed Molecular Distribution of Red and Blue Clouds \label{3-2}}
\begin{figure}[htbp]
 \begin{center}
  \includegraphics[width=14cm]{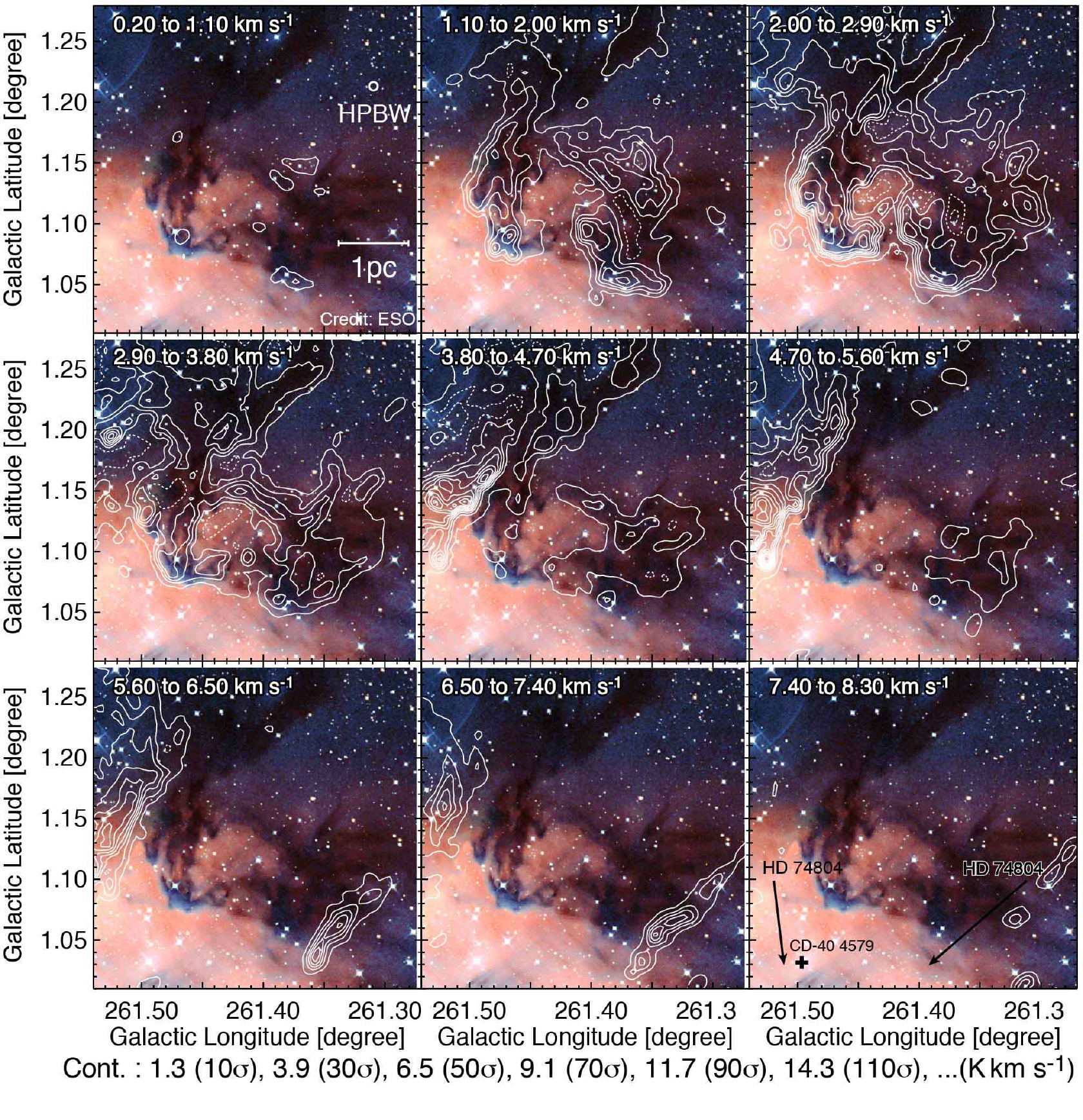} 
 \end{center}
\caption{Velocity channel distribution of $^{12}$CO ($J$ = 3--2) obtained by the ASTE telescope superposed on the optical image of Figure~2a (Credit: eso) as contours. Depression contours are indicated as dotted curves. The direction of the early B star, HD~74804, is indicated by black arrows in the bottom--right panel. The cross indicates the position of CD$-$40~4579 for reference.}\label{ch32}
\end{figure}

Figure~5 shows velocity--channel distributions of the $^{12}$CO ($J$ = 3--2) emission obtained with ASTE superposed on three optical color composite images in Figure~2a as contours. The direction of early B star, HD~74804 is indicated by black arrows in the bottom--right panel. The $J$ = 3--2 data with a finer spatial resolution clearly show elongated filamentary molecular clouds in the Blue cloud. In a velocity range from 0.2 to 5.6 km~s$^{-1}$, the Blue cloud corresponds to the optical dark cloud. The shape of the Blue cloud indicates that the cloud is influenced by the B star(s). Dark features in $l$ = 261\fdg45 -- 261\fdg50, $b$ = 1\fdg07 -- 1\fdg15 are influenced by the UV radiation from the B star(s) and exhibit distribution similar to {\it the pillars of creation} in the Eagle nebula (M16). In a velocity range from 3.8 to 7.4 km~s$^{-1}$, the Red cloud North is distributed along the northeastern edge of the Blue cloud. In a velocity range from 5.6 to 8.3 km~s$^{-1}$, the Red cloud North is distributed along the southwestern edge of the Blue cloud.

\subsection{Intensity Ratio of $^{12}$CO$J$ = 3--2 / 2--1}
\begin{figure}[htbp]
 \begin{center}
  \includegraphics[width=14cm]{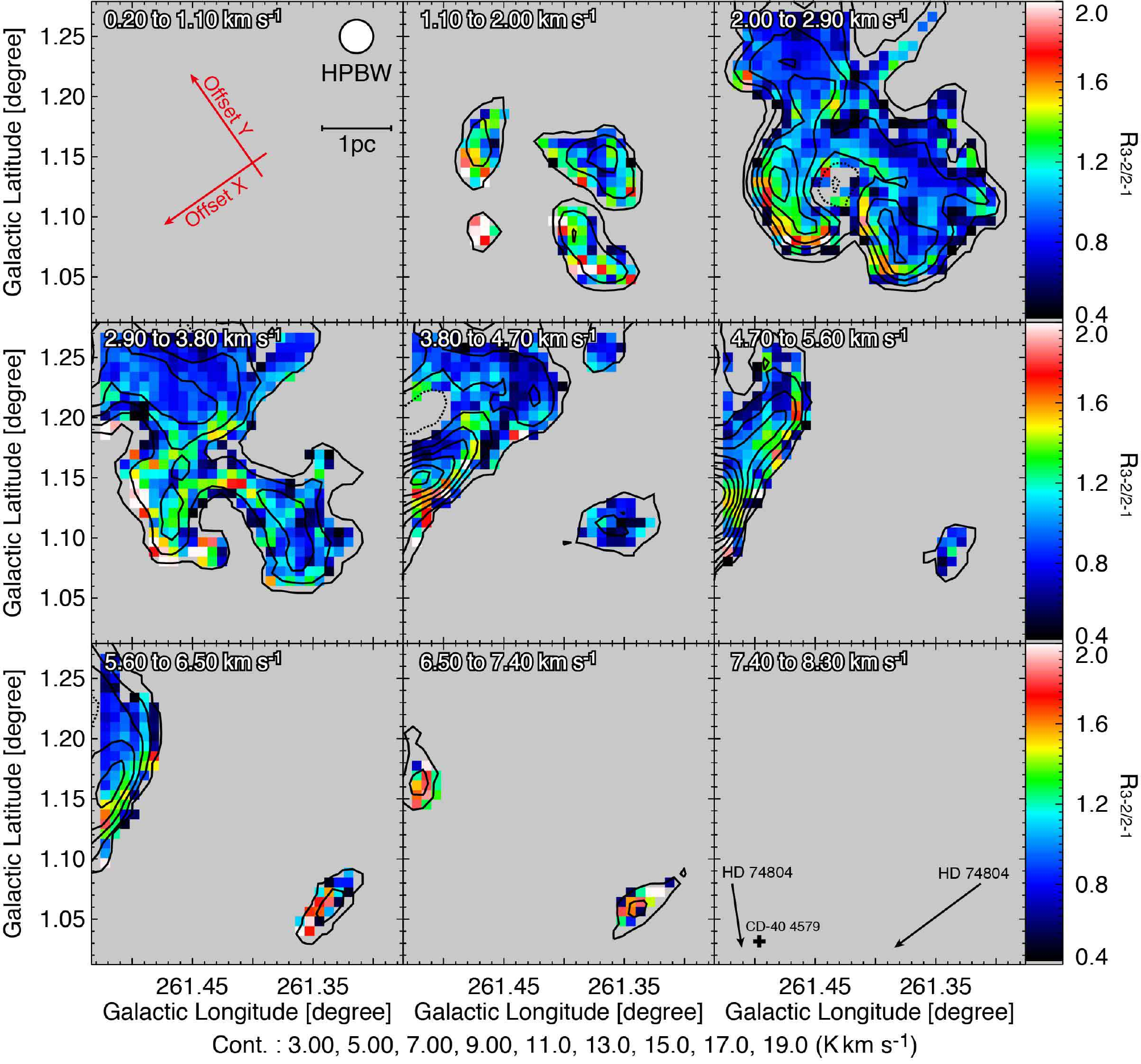} 
 \end{center}
\caption{Velocity channel distribution of $R_{3-2/2-1}$. Contours show distribution of integrated intensity of $^{12}$CO ($J$ = 2--1). Depression contours are indicated as dotted curves. The direction of the exciting early B star, HD~74804 is indicated by the directions of black arrows in the right bottom panel. The black cross indicates the position of CD$-$40~4579 for reference. The red arrows in the top--left panel indicate the offset X--Y coordinate makes an angle of 35 degrees to the galactic coordinate with an origin at ($l$, $b$) = (261\fdg400, 1\fdg145) used in Figure 7.}\label{ratio}
\end{figure}

Figure~6 shows velocity--channel distributions of an intensity ratio $R_{3-2/2-1}$ of the $^{12}$CO ($J$ = 3--2) emission to the $^{12}$CO ($J$ = 2--1) emission. We \textcolor{black}{convolved both the $^{12}$CO ($J$ = 2--1) and $^{12}$CO ($J$ = 3--2) data with Gaussian functions to have the same HPBW, 103$\arcsec$ and } used the pixels with intensities higher than eight sigmas for the both data. Contours show the integrated intensity of the $^{12}$CO ($J$ = 2--1) emission. The direction of early B star, HD~74804 is indicated by black arrows in the bottom--right panel.

We use the published data in order to estimate the typical line intensity ratio $R_{3-2/2-1}$ in the Galactic disk, by choosing regions where no local extra heating is working. Since there is no extensive survey in the $J$ = 3--2 emission, we adopt individual objects which were observed in the $J$ = 3--2, 2--1, and 1--0 transitions of $^{12}$CO. One of the objects selected is RCW~38 at a distance of 1.7 kpc which includes an H{\sc ii} region observed at 0.3 pc resolution in the $J$ = 3--2 and $J$ = 1--0 transitions (their intensity ratio is defined as $R_{3-2/1-0}$) by Fukui et al. (2016). Another object is M20, an H{\sc ii} region at a distance of 1.7 kpc, and is observed in the $J$ = 2--1 and 1--0 transitions (their intensity ratio is defined as $R_{2-1/1-0}$) at 1 pc resolution by Torii et al. (2011). We find that $R_{3-2/2-1}$ $\simeq$ 0.4 in RCW~38 and $R_{2-1/1-0}$ $\simeq$ 0.5 in M20 and that the two ratios are fairly uniform in regions over a $\sim$10 pc extent where no extra heating is noticed. We consider that the two ratios are typical to the disk clouds without an extra heat source and estimate a typical ratio $R_{3-2/2-1}$ of 0.8 by taking a ratio of the two values.
\textcolor{black}{Unlike $^{12}$CO ($J$ = 2--1), distribution of $^{12}$CO ($J$ = 3--2) emission observed with $\sim$4 times finer angular resolution shows filamentary structures (See Figure 5). It is possible that the obtained $R_{3-2/2-1}$ gives lower limit as it is smoothed down to the coarser resolution of $J$ = 2--1.
However, this effect should not be significant because $R_{3-2/2-1}$ only at the surfaces of molecular clouds facing the B star show higher values due to the irradiation of UV radiation while the other regions show lower values which is almost consistent with the galactic typical one (0.8).}

In a velocity range from 1.1 to 3.8 km~s$^{-1}$, the Blue cloud exhibits a gradient in $R_{3-2/2-1}$ toward B star(s). The inner part of the Blue cloud facing to the H{\sc ii} region shows $R_{3-2/2-1}$ higher than 1.0 up to 2.0, whereas the outer part of the cloud shows $R_{3-2/2-1}$ lower than 0.8. This suggests that the Blue cloud is associated with the H{\sc ii} region and is heated up radiatively. In a velocity range from 3.8 to 7.4 km~s$^{-1}$, the Red cloud North shows a gradient of $R_{3-2/2-1}$ toward its sharp edge and high $R_{3-2/2-1}$ over 1.5 are seen on the edge. The edge corresponds to SFO 58 and its exciting star is early B star HD~74804 (Urquhart et al. 2006). In a velocity range from 3.8 to 5.6 km~s$^{-1}$, part of the Blue cloud at ($l$, $b$) $\sim$ (261\fdg35, 1\fdg10), a ratio of typical value for the Galactic disk and is not significantly affected by the B stars. The Blue cloud and part of the Red cloud North show a high $R_{3-2/2-1}$ along the southern edge in Figure~6. This may be explained as due to heating by HD~74804, while some contribution of CD$-$40 4579 is not excluded.
In a velocity range from 5.6 to 7.4 km~s$^{-1}$, part of the Red cloud North at ($l$, $b$) $\sim$ (261\fdg35, 1\fdg05), shows a gradient of a ratio toward B stars with significantly high values from 0.9 to 2.0 in the region facing to the B star HD~74804.

\subsection{Physical Parameters of the Molecular Clouds}
In this subsection, we derive physical parameters such as molecular mass and column density for each cloud. Molecular mass {\it M} in $M_{\odot}$ is estimated from Equations (1).

\begin{equation}
  M = \mu m_p \sum_{i} [d^2 \Omega N_i($H$_2)]
\end{equation}
where $\mu$, $m_p$, $d$, $\Omega$ and $N_i($H$_2)$ are mean molecular weight, mass of hydrogen, distance, solid angle of a pixel, and column density of molecular hydrogen for i-th pixel, respectively. We use 20 $\%$ for abundance of helium corresponding $\mu$ = 2.8 and $d$ = 1.0 kpc. Column densities of molecular hydrogen for each cloud is estimated by Equation (2).

\begin{equation}
  N($H$_2) = X \times W(^{12}$CO$)
\end{equation}
where $W$($^{12}$CO) is integrated intensity of $^{12}$CO ($J$ = 1--0) and $X$ is an empirical conversion factor from $W$($^{12}$CO) to {\it N}(H$_2$). We adopt $X$ = 1.0 $\times$ 10$^{20}$ cm$^{-2}$ (K km s$^{-1}$), which is derived by Okamoto et al. (2017) by using the $Planck$ dust emission data toward the Perseus molecular cloud. From these equations, molecular masses for the Blue cloud, the Red cloud North and the Red cloud South are derived to be 360 $M_\odot$, 900 $M_\odot$, and 300 $M_\odot$ and maximum column densities are $\sim$ 2$\times$ 10$^{21}$ cm$^{-2}$, $\sim$ 6$\times$ 10$^{21}$ cm$^{-2}$, and $\sim$ 6$\times$ 10$^{21}$ cm$^{-2}$, respectively. Detailed physical parameters for each cloud are summarized in Table 1.

\begin{table*}[h]
\tbl{Physical parameters of the molecular clouds}{%
\begin{tabular}{lcccc} 
\hline\noalign{\vskip3pt}
Cloud name & $V_{LSR}$ [km s$^{-1}$] & Peak Velocity [km~s$^{-1}$] & Molecular mass [$M_\odot$] & $N$(H$_2$)  [cm$^{-2}$]   \\  
\hline\noalign{\vskip3pt} 
Blue cloud & 1.4 -- 3.8 & 2.8 & 360 & 2 $\times$ 10$^{21}$\\
Red cloud North & 2.6 -- 6.8 & 5.2 & 900 & 6$\times$ 10$^{21}$\\
Red cloud South & 2.6 -- 6.8 & 5.4 & 300 & 6$\times$ 10$^{21}$\\
SL2 cloud & 5.6 -- 7.4 & 6.3 & 60 & 1$\times$ 10$^{21}$\\
\hline\noalign{\vskip3pt} 
\end{tabular}}
\label{table:physpara}
\begin{tabnote}
\hangindent6pt\noindent
Note. --- Col.1: Cloud's name. Col.2\textcolor{black}{:} Velocity span derived from $^{12}$CO ($J$ = 2--1) dataset. Col.3\textcolor{black}{:} Peak velocity derived from a gaussian fit of an average profile of each clouds in $^{12}$CO ($J$ = 2--1). Col.4\textcolor{black}{:} Molecular mass. Col.5\textcolor{black}{:} Maximum molecular column density $N$(H$_2$) toward each cloud.\\
\end{tabnote}
\end{table*}

\section{Discussion}
\subsection{Molecular Gas Distribution}
We summarize the present observational results as follows;

\begin{enumerate}
 \item Molecular clouds toward RCW~32 are distributed in four different components, the Blue cloud, the Red cloud North, the Red cloud South, and the SL2 cloud.
 \item The Red cloud South, corresponding to SFO 57, has $V_{LSR}$ = 2.6 -- 6.8 km~s$^{-1}$ and 300 $M_{\odot}$. The Red cloud North, corresponding to SFO 58, has $V_{LSR}$ = 2.6 -- 6.8 km~s$^{-1}$ and 900 $M_{\odot}$. The Blue cloud with $V_{LSR}$ = 1.4 -- 3.8 km~s$^{-1}$ and 360 $M_{\odot}$ has a counterpart in the \textcolor{black}{optically} dark features in the northwest of RCW 32. The SL2 cloud at 5.6 -- 7.4 km~s$^{-1}$ corresponds to the central dark lane SL2.
 \item The Blue cloud is on the near side of RCW~32 while the Red clouds North and South are located inside or on the far side of RCW~32.
 \item We found complementary spatial distribution between the Red cloud North and the Blue cloud which suggests their physical interaction, while these two clouds have different radial velocity by $\sim$2.4 km~s$^{-1}$. 
 \item Both the Red and Blue clouds have regions with a ratio $R_{3-2/2-1}$ over 1.2 which is substantially higher as compared with the typical value in Galactic disk clouds ($\sim$ 0.8). 
\end{enumerate}

\subsection{Parent Cloud(s) of the Star Cluster Cr 197}
Figure~2a suggests that RCW~32 is not a young H{\sc ii} region because the central B stars are not associated with the parent clouds within 3 -- 4 pc except \textcolor{black}{for} SFO~57 and SFO~58. It is likely that the molecular gas has been dispersed by ionization and winds from HD74804. We derived the electron density toward the peak of radio continuum emission by the 4850 MHz Parkes-Mit-NRAO (PMN) survey (Griffith $\&$ Wright 1993). The electron density is estimated to be $\sim$ 20 cm$^{-3}$ by assuming an electron temperature of 8000 K and a path length as 6.5 pc, the same with the size of the H{\sc ii} region. According to Israel (1976b) this electron density corresponds to a classical H{\sc ii} region which means a relatively evolved H{\sc ii} region. Moreover, Tremblin et al. (2014) carried out 3-dimensional hydrodynamical simulations of expanding H{\sc ii} regions and showed that it takes more than 1 Myr for an H{\sc ii} region to expand to a radius of 3 pc. This is consistent with the interpretation above.
 There is no work which addresses the formation of the exciting star in RCW~32. The present results on detailed molecular distribution opened a possibility to explore the formation mechanism. We found that the two Red clouds in North and South have similar ranges of radial velocities, peak velocities, and column densities. Furthermore, it is probable that both of them are ionized by the same source (HD~74804) and form bright rims toward the side close to the ionization source (see Figure~3 and Table~\ref{table:physpara}).
This suggests that there was a single large cloud which was distributed between the Red clouds North and South before the B star \textcolor{black}{formation}. SL2 is an outstanding dark feature in optical images as a dark feature (e.g., Figure 2a), whereas it is not bright at the infrared wavelengths (Figure 4b). This suggests that SL2 is not located close to RCW~32 and is not irradiated by the H{\sc ii} region. Conversely, Figure 4b shows that SFO~57 and SFO~58 are emitting infrared radiation and they are likely associated with RCW~32.

\subsection{Possible Origin of High Mass Stars}
First, we examine a possibility of a simple model that the exciting star was formed in a natal cloud in the past and the expanding motion was produced by the stellar wind which caused the observed velocity span between the Red and Blue clouds.

Total mass of the associated clouds is calculated to be $\sim$1600 $M_\odot$. An angle of the cloud relative motion to the line of sight is tentatively assumed to be 45 degrees and then the corrected relative velocity between the Blue and the Red clouds is calculated as 4 (= 2.6 $\times \sqrt{2}$ ) km~s$^{-1}$. The required mass to gravitationally bind the two clouds with a relative velocity of 4 km~s$^{-1}$ and a radius of 4.5 pc is calculated as 1.6 $\times$ 10$^4$ $M_\odot$. This value is an order of magnitude higher than the mass of clouds and the cluster itself, and thus the Red and Blue clouds are gravitationally unbound. Then, the mass of the two northern clouds (the Red cloud North and the Blue cloud) is 900 $M_{\odot}$ and 360 $M_\odot,$ respectively and the required momentum to move these two clouds with a relative velocity of 4 km s$^{-1}$ is calculated as $\sim$1 $\times$ 10$^3$ $M_\odot$ km~s$^{-1}$. This values is almost the same with the analytical solution of momentum of O type star's stellar wind ($\sim$2 $\times$10$^3$ $M_\odot$ km s$^{-1}$; c.f., Abbott 1982a, b).
The Blue cloud is separated from HD~74804 and the projected angle subtended by the Blue cloud to the star is $\sim$50 degrees. This angle is converted into the solid angle by assuming that the Blue cloud has a depth similar to its width in the sky. We thus estimate the solid angle subtended by the star to the Blue cloud is $\sim$0.6 str, indicating that $\sim$5 $\%$ of the stellar wind momentum, 100 $M_{\odot}$ km~s$^{-1}$, is available to accelerate the Blue cloud, which is too small to explain the velocity shift. We therefore conclude that the stellar winds are not able to cause the velocity shift of the Blue cloud, and that the velocity span is hardly explained as due to the stellar wind. 

Next, we examine the effect of an H{\sc ii} region on the cloud dynamics. We used the model of gas acceleration by an early B star in hydrodynamical numerical simulations by Hosokawa $\&$ Inutsuka (2007). We chose their model CNM-S12 and found the effects of the B star on the molecular gas as a velocity shift of $\sim$ 2 km s$^{-1}$ and a thickness of the compressed layer of less than 1 pc (Figure 5 in the paper). The results indicate a typical outcome of acceleration by the ultraviolet photons. 
We developed discussion that the velocity field accelerated by a B star should show a velocity shift of a few km~s$^{-1}$ at places closer to the stars. Such a trend is not found in the present data (see Figures 3 and 5). Accordingly, we reached the same conclusion that the gas motion is not significantly affected by HD~74804. The blue and red shifts of the clouds are likely preexistent, and are not to due the stellar acceleration. We suggest that the accelerated gas is possibly ionized already. It is also probable that the expanding H{\sc ii} gas easily escapes in the highly inhomogeneous molecular distribution as the so-called Champagne flow (Tenorio-Tagle 1979).
We examine the possibility of a cloud-cloud collision as an alternative model in the following.

\subsection{Cloud-Cloud Collision as a Triggering Mechanism of High-Mass Star Formation}
It was discussed that cloud--cloud collision is a triggering mechanism of star burst activities in colliding galaxies (e.g.,Young et al.1986). Recent studies indicate that cloud--cloud collision phenomena frequently take place not only in interacting galaxies but also among molecular clouds in the Galaxy. Dobbs et al. (2015) carried out hydrodynamical numerical simulations of isolated galaxies and found that collisions of molecular clouds having a size larger than 10 pc occur every 8 -- 10 Myr. The mean free time between collisions is sufficiently shorter than a life time of molecular clouds in a galaxy, $\sim$20 Myrs (Fukui et al. 2008;  et al. 2009; Fukui \& Kawamura 2010). Theoretically, Inoue $\&$ Fukui (2013) demonstrated that cloud--cloud collision realizes a large mass accretion rate of 10$^{-4}$ -- 10$^{-3}$ $M_\odot$yr$^{-1}$ through three-dimensional magnetohydrodynamics (MHD) simulations, and that the high mass accretion rate leads to form massive, gravitationally bound molecular cloud cores, precursors of high mass stars. Habe $\&$ Ohta (1992) and Anathpindika (2009b) showed that a collision between two molecular clouds with supersonic relative velocities generates a compressed layer at the collisional front, and that enhanced turbulence in the layer increases an effective sound speed and then increases the Jeans mass. Recent observations presented evidence on cloud--cloud collision in super star clusters (e.g., Westerlund 2, Furukawa et al. 2009; NGC3603, Fukui et al. 2013) and Galactic open clusters / H{\sc ii} regions (e.g., M20; Torii et al. 2011, RCW~120; Torii et al. 2015). These authors discovered two natal molecular clouds with various signatures of collision toward the young high mass stars. Three H{\sc ii} regions RCW~34, RCW~36, and RCW~38 in Vela C are also proposed as sites where cloud--cloud collision occurred and triggered the formation of O / early B stars (Hayashi et al. 2017; Sano et al. 2017; Fukui et al. 2016). These authors discovered that the two natal clouds exhibit complementary spatial distribution, an observational signature of collision. It is therefore worthwhile to explore if cloud--cloud collision is a dominant mechanism in the VMR, and RCW~32 is an obvious object to test if cloud--cloud collision is operating.

\subsection{Cloud-Cloud Collision in RCW~32}
\begin{figure}[htbp]
 \begin{center}
  \includegraphics[width=8cm]{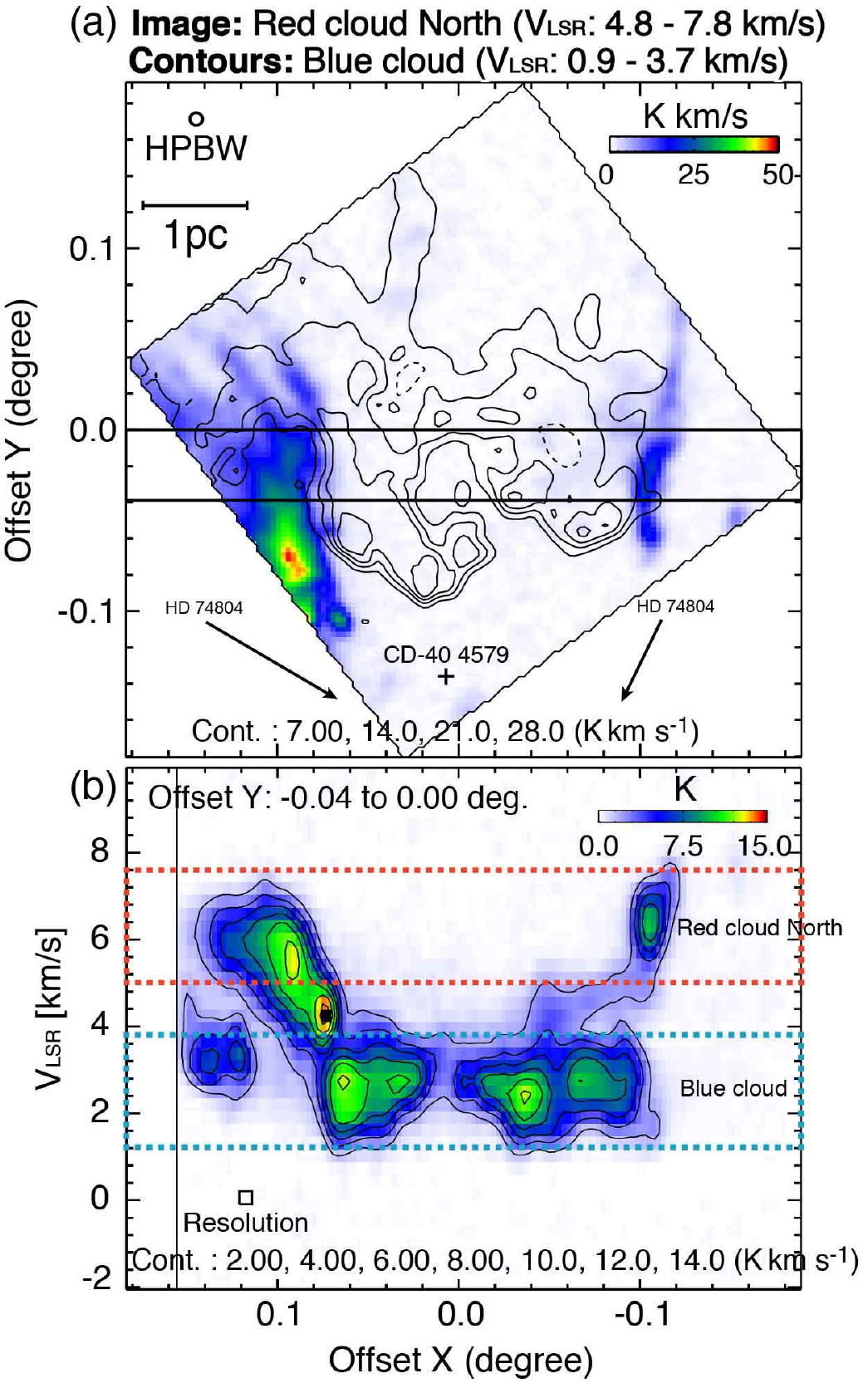} 
 \end{center}
\caption{(a) Integrated intensity distribution of the Red cloud North (image) and the Blue cloud (contours) in $^{12}$CO ($J$ = 3--2) in Offset X and Y coordinate. Depression contours are indicated as dotted curves. The integrated velocity range for each cloud is shown in the figure. The direction of another early B star, HD~74804 is indicated by black arrows. The black cross indicates the position of CD$-$40~4579 for reference. (b) Position--Velocity diagram of $^{12}$CO ($J$ = 3--2) in Offset X and Y coordinate. The integrated offset Y range is shown as the black box in the panel (a).}\label{comple}
\end{figure}

\subsubsection{The Complementary Distribution and the Bridge Features}
The Blue cloud is located in the north and the Red clouds have two features in the north and south, i.e., the Red clouds North and South. The two velocity components are physically associated with HD~74804 and are dispersed within 3 -- 4 pc of the stars probably due to the ionization and stellar winds.
Figure~7 shows a spatial distribution and position--velocity diagram of $^{12}$CO ($J$ = 3--2) toward the Blue cloud obtained with ASTE. The coordinate in the figure is taken along the direction of the interface between the Blue cloud and the Red cloud North. We defined the Offset X and Y axes through rotating the galactic coordinate system counterclockwise by 35 degrees centered on ($l$, $b$) = (261\fdg400, 1\fdg145) as shown in Figure 6.
Figure~7a shows integrated intensity distribution of the Red cloud North (image) and the Blue cloud (contours) in $^{12}$CO ($J$ = 3--2), and indicates that two clouds show complementary distribution, where the Blue cloud fits the edge of the Red cloud North. Figure~7b shows a position(offset X)-velocity diagram toward the Blue cloud and the Red cloud North, and indicates that the two clouds have connecting bridge features at 4 -- 5 km~s$^{-1}$ toward the both edges of the Blue cloud at X = -0.1 -- -0.05 degrees and X = 0.1 degrees.
The integrated range in offset Y is shown by the black box in Figure 7a. The line intensity ratio and the BRCs provide observational signs of physical association and interaction with HD~74804. The complementary distribution and the bridge feature are observational signatures of cloud-cloud collision (Torii et al. 2017; Fukui et al. 2016)

The direction of the edges of the two clouds in Figure 7\textcolor{black}{a} seems to be parallel with each other. A possible interpretation of these edges is that the clouds are dynamically affected by HD~74804 instead of the collisional interaction. It is however not supported by that the directions of their edges are not coincident with the direction of HD~74804 which is shown by two arrows in Figure 7a. We thus consider only the collisional interaction in the present paper.
The filamentary molecular distributions are observed in some places where cloud-cloud-collision takes place (N159 in LMC; Fukui et al. 2015, RCW 79; Ohama et al. 2017).
Furthermore, Inoue \& Fukui (2013) performed 3D MHD simulations of cloud-cloud-collision and found that dense filamentary molecular clouds are formed through cloud-cloud collision by an anisotropic compression along magnetic field lines. The filamentary cloud distributions in Figure 5 may be explained in terms of the collisional interaction which causes shearing motion between the clouds, whereas the effect of the expanding H{\sc ii} region may be working also to form part of the cloud distribution.

\subsubsection{A Possible Scenario of Cloud-Cloud Collision}

\begin{figure}[htbp]
 \begin{center}
  \includegraphics[width=14cm]{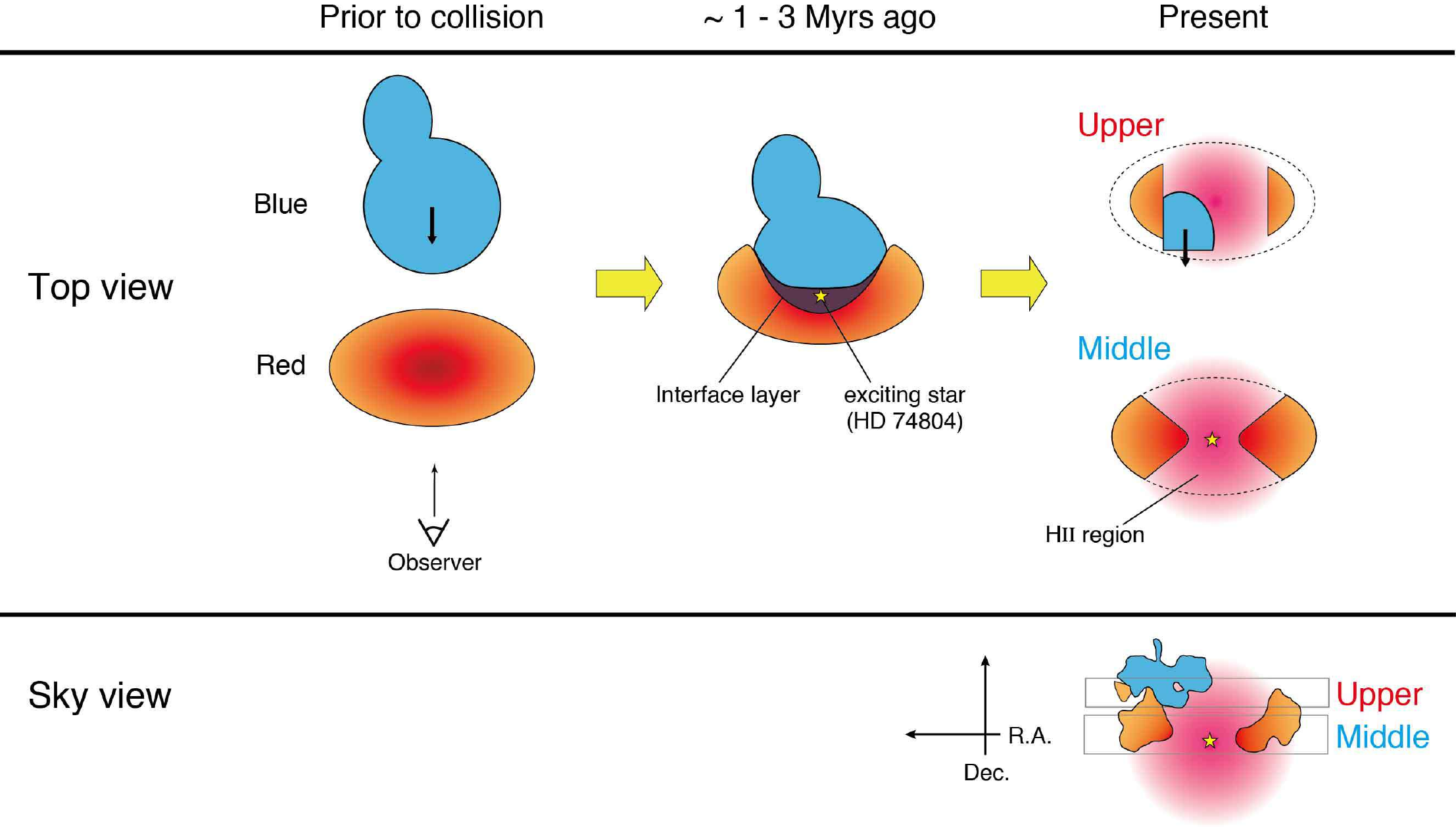} 
 \end{center}
\caption{Schematic image of cloud--cloud collision in RCW~32. For the details see text.}\label{schematic}
\end{figure}

A possible configuration of cloud--cloud collision is schematically shown in Figure 8. The Blue cloud collided with the Red cloud, where the Blue cloud moves from the North toward us.
An estimate of the collision time scale is possible by adopting a cloud size of 10 -- 20 pc and a relative velocity of 4 km~s$^{-1}$, and we obtain roughly $\sim$2 Myrs as the collision duration from a ratio of the size and velocity, which gives roughly a timescale of the collisional interaction. This collision triggered formation of HD~74804 in their overlapped region. Since the molecular gas close to the B stars are already almost fully dispersed most likely due to the ionization and winds, we assume that the collision terminated $\sim$1 Myr ago, which favors the cluster age  $\sim$1 Myr (Pettersson $\&$ Reipurth 1994) rather than $\sim$5 $\pm$ 4 Myrs (Bonatto $\&$ Bica 2010). 
A possible scenario is as follows; $\sim$3 Myr ago the collision started and $\sim$1 Myr ago the collision ended.  In the collision, the Blue cloud created a cavity in the Red cloud and triggered the formation of HD~74804 in the interface layer between the clouds 1 Myr ago (Figure~8). 
High mass star formation triggered in cloud--cloud collision is supported by the MHD numerical simulation of Inoue \& Fukui (2013). HD~74804 dispersed the natal clouds within 3 -- 4 pc of these stars. The location of the Blue cloud is on the near side of the H{\sc ii} region as evidenced by the dark lanes. The present shapes of the Red and Blue clouds may still hold original distributions prior to the collision.

The two signatures above, the complementary distribution and the bridges characteristic to cloud--cloud collision suggests that collision took place in RCW~32. This does not directly support that the star formation was triggered by the cloud--cloud collision because the natal clouds are strongly dispersed within 3 -- 4 pc of HD~74804, making it difficult to assess the natal gas properties. Therefore, RCW~32 is a case where triggered high mass star formation is not directly confirmed based on molecular clouds due to the cloud dispersal. This differs from a case RCW~38 where collisional triggering is firmly supported by the current cloud distribution thanks to a very young cluster age 0.1 Myr (Fukui et al. 2016).

According to the previous samples of triggered high mass star formation, the molecular column density is as high as 10$^{22}$ cm$^{-2}$ for single O / early B star formation by collisional triggering (e.g., Torii et al. 2011; Fukui et al. 2016). The cloud column density of the Red cloud in RCW~32 is 6 $\times$ 10$^{21}$ cm$^{-2}$, close to this column density, and is not in contradiction with the early B star formation. It requires further investigations to test if the cluster Cr 197 was also formed by the cloud-cloud collision. In the other objects including RCW38 and M42 (Fukui et al. 2016; 2017), it was suggested that a cluster was formed prior to the collision which formed the O / early B stars over a longer time scale than the collisional duration. The star formation history in Cr 197 has not been studied into detail and additional observational studies of cluster members are desirable.

\section{Conclusions}
We observed molecular clouds toward RCW~32 in $^{12}$CO ($J$ = 1--0, 2--1, and 3--2) and $^{13}$CO ($J$ = 1--0 and 2--1) with the NANTEN2 and ASTE telescopes. The results lend support for the formation of Cr 197 including the exciting star of it through cloud--cloud collision. We summarize conclusions as follows.

\begin{enumerate}
 \item We revealed that three molecular clouds are associated with RCW 32 in CO observations. The Red cloud South corresponds to SFO 57 has $V_{LSR}$ = 2.6 -- 6.8 km~s$^{-1}$ and 300 $M_\odot$, and the Red cloud North corresponds to SFO 58 has $V_{LSR}$ = 2.6 -- 6.8 km~s$^{-1}$ and 900 $M_\odot$. The Blue cloud has a counterpart to the optical dark cloud in the north west of RCW~32 with $V_{LSR}$= 1.4 -- 3.8 km~s$^{-1}$ and 360 $M_\odot$.
  \item Both the Red and the Blue clouds are strongly ionized by UV radiation from the exciting star(s) and show significantly high values of $R_{3-2/2-1}$ than typical galactic disk clouds.
 \item We found the obvious complementary spatial distribution and connecting bridge features between the Red and the Blue cloud. These signatures are consistent with the collisional interaction between the clouds. We argued that the cloud distribution and kinematics are hard to explain by a simple expanding model. We present a hypothesis that a collision between the Red and Blue clouds took place from $\sim$3 Myrs ago in a duration of $\sim$2 Myrs and formed the exciting stars and the cluster whose age is $\sim$1 Myr. The natal cloud dispersal is significant and we see a cavity of 4 -- 5 pc radius centered on the exciting star. By including the previous cases of cloud-cloud collision in the VMR, e.g., , RCW~34,  RCW~36, and RCW~38, an increasing number of H{\sc ii} region formed by cloud-cloud collision is perhaps due to the high molecular density in the region.
\end{enumerate}

\begin{ack}
We thank an anonymous referee for very helpful comments that improved the manuscript.
NANTEN2 is an international collaboration of 10 universities: Nagoya University, Osaka Prefecture University, University of Bonn, University of Cologne, Seoul National University, University of Chile, University of New South Wales, Macquarie University, University of Sydney, and University of ETH Zurich. The ASTE telescope is operated by NAOJ. The Southern H-Alpha Sky Survey Atlas (SHASSA), which is supported by the National Science Foundation.
This publication makes use of data products from the Wide-field Infrared Survey Explorer, which is a joint project of the University of California, Los Angeles, and the Jet Propulsion Laboratory/California Institute of Technology, funded by the National Aeronautics and Space Administration.
This work was financially supported by Grants-in-Aid for Scientific Research (KAKENHI) of the Japanese society for the Promotion of Science (JSPS, grant No. 15H05694). This work also was supported by “Building of Consortia for the Development of Human Resources in Science and Technology” of Ministry of Education, Culture, Sports, Science and Technology (MEXT, grant No. 01-M1-0305). We also acknowledge to Daiki Kurita and Ryuji Okamoto for contributions on the observations of $^{12}$CO ($J$ = 3--2) data.
\end{ack}



\end{document}